\documentclass[%
 reprint,
superscriptaddress,
 amsmath,amssymb,
 aps,
]{revtex4-1}
\usepackage{physics}
\usepackage{graphicx}% Include figure files
\usepackage{dcolumn}% Align table columns on decimal point
\usepackage{bm}% bold math
\usepackage{multirow}
\usepackage{threeparttable}
\usepackage{array}
\usepackage{textcomp}
\usepackage{float}
\begin{document}

\title{ Probing the quadrupole transition strength of $^{15}$C via deuteron inelastic scattering } % Force line breaks with \\

\author{J.~Chen}
 \email{jiechenphysics@gmail.com}
\affiliation{Physics Division, Argonne National Laboratory, Argonne, Illinois 60439, USA}

\author{B.~P.~Kay}
\affiliation{Physics Division, Argonne National Laboratory, Argonne, Illinois 60439, USA}

\author{T.~L.~Tang}
\email{Present address: Department of Physics, Florida State University, Tallahassee, Florida 32306, USA}
\affiliation{Physics Division, Argonne National Laboratory, Argonne, Illinois 60439, USA}

\author{I.~A.~Tolstukhin}
\affiliation{Physics Division, Argonne National Laboratory, Argonne, Illinois 60439, USA}

\author{C.~R.~Hoffman}
\affiliation{Physics Division, Argonne National Laboratory, Argonne, Illinois 60439, USA}

\author{H.~Li}
\affiliation{Institute of Modern Physics, Chinese Academy of Sciences, Lanzhou 730000, China}

\author{P.~Yin}
\affiliation{Institute of Modern Physics, Chinese Academy of Sciences, Lanzhou 730000, China}

\author{X.~Zhao}
\affiliation{Institute of Modern Physics, Chinese Academy of Sciences, Lanzhou 730000, China}

\author{P.~Maris}
\affiliation{Department of Physics and Astronomy, Iowa State University, Ames, Iowa 50011, USA}

\author{J.~P.~Vary}
\affiliation{Department of Physics and Astronomy, Iowa State University, Ames, Iowa 50011, USA}

\author{G.~Li}
\affiliation{%
 School of Physics and State Key Laboratory of Nuclear Physics and Technology, Peking University, Beijing 100871, China}%

\author{J.~L.~Lou}
\affiliation{%
School of Physics and State Key Laboratory of Nuclear Physics and Technology, Peking University, Beijing 100871, China}%

\author{M.~L.~Avila}
\affiliation{Physics Division, Argonne National Laboratory, Argonne, Illinois 60439, USA}

\author{Y.~Ayyad}
\affiliation{%
IGFAE, Universidade de Santiago de Compostela, E-15782, Santiago de Compostela, Spain}

\author{S.~Bennett}
\affiliation{Department of Physics and Astronomy, University of Manchester, M13 9PL Manchester, United Kingdom}

\author{D.~Bazin}
\affiliation{National Superconducting Cyclotron Laboratory, Michigan State University, 640 S Shaw Ln, East Lansing, Michigan 48824, USA}

\author{J.~A.~Clark}
\affiliation{Physics Division, Argonne National Laboratory, Argonne, Illinois 60439, USA}

\author{S.~J.~Freeman}
\affiliation{Department of Physics and Astronomy, University of Manchester, M13 9PL Manchester, United Kingdom}
\affiliation{EP Department, CERN, Geneva CH-1211, Switzerland}

\author{H.~Jayatissa}
\affiliation{Physics Division, Argonne National Laboratory, Argonne, Illinois 60439, USA}

\author{C.~Müller-Gatermann}
\affiliation{Physics Division, Argonne National Laboratory, Argonne, Illinois 60439, USA}

\author{A.~Munoz-Ramos}
\affiliation{%
IGFAE, Universidade de Santiago de Compostela, E-15782, Santiago de Compostela, Spain}

\author{D.~Santiago-Gonzalez}
\affiliation{Physics Division, Argonne National Laboratory, Argonne, Illinois 60439, USA}

\author{D.~K.~Sharp}
\affiliation{Department of Physics and Astronomy, University of Manchester, M13 9PL Manchester, United Kingdom}

\author{A.~H.~Wuosmaa}
\affiliation{Department of Physics, University of Connecticut, Storrs Connecticut 06269, USA}

\author{C.~X.~Yuan}
\affiliation{%
Sino-French Institute of Nuclear Engineering and Technology, Sun Yat-Sen University, Zhuhai 519082, China
}%

%\footnote[1]{Present address: Department of Physics, Florida State University, Tallahassee, Florida 32306, USA }

\date{\today}

\begin{abstract}
Deuteron elastic scattering from $^{15}$C and inelastic scattering reactions to the first excited state of $^{15}$C were studied using a radioactive beam of $^{15}$C in inverse kinematics. The scattered deuterons were  measured using HELIOS. The elastic scattering differential cross sections were analyzed using the optical model. A matter deformation length $\delta_d=1.04(11)$\,fm has been extracted from the differential cross sections of inelastic scattering to the first excited state. The ratio of neutron and proton matrix elements $M_n/M_p = 3.6(4)$ has been determined from this quadrupole transition. Neutron effective charges and core-polarization parameters of $^{15}$C were determined and discussed. Results from {\it ab-initio} no-core configuration interaction calculations were also compared with the experimental observations. This result supports a moderate core decoupling effect of the valence neutron in $^{15}$C similarly to its isotone $^{17}$O, in line with the interpretation of other neutron-rich carbon isotopes. 
\end{abstract}

\maketitle

\section{Introduction}

Halo nuclei have been extensively studied in the past decades owing to the availability of radioactive beams~\cite{Tanihata1985, TANIHATA,Hansen1987}. A one-neutron halo nucleus is composed of one weakly-bound valence neutron coupled to the core and has a large matter radius induced by the spatially extended character of the valence neutron wave function. This valence neutron usually occupies a single-particle orbital with a low angular momentum, preferably the $2s_{1/2}$ orbital, for example in $^{11}$Be, $^{15}$C and $^{19}$C. The weakly bound low-lying states will couple to the continuum, which may impact the reaction mechanism, and the interplay between core and the valence nucleon may  change the differential cross sections of the elastic and inelastic scattering~\cite{CHEN2016,CHEN2016Mar}.  %Particularly, isoscalar soft dipole resonances have been observed in $^{11}$Li in recent years.

 Valence-neutron decoupling from the inert core in the neutron-rich nuclei is also a prominent phenomenon.  The reduced effective charge  is a measure of the magnitude of core polarization induced by the valence neutrons~\cite{Raimondi, Sagawa}. Another interesting feature related to the decoupling is an enhanced soft dipole excitation mode, with   enhanced direct breakup cross sections at low excitation energy found in  $^{11}$Be~\cite{Fukuda}, $^{15}$C~\cite{Datta}, $^{19}$B~\cite{Cook}, $^{31}$Ne~\cite{Nakamura2009} and $^{11}$Li~\cite{Nakamura2006}. The degree of coupling can also be described by the difference of the neutron and proton contributions to the quadrupole transition. It is determined by the ratio of neutron and proton quadrupole matrix elements $M_n/M_p = N\delta_n/(Z\delta_p)$, where $\delta_n$ and $\delta_p$ are the neutron and proton deformation lengths, respectively~\cite{Bernstein1983}. This ratio is usually close to $N/Z$ for nuclei where the neutrons and protons are strongly coupled, so their contributions are similar. However, for nuclei with large $N/Z$ ratios, the large difference in the proton and neutron Fermi energy  may lead to the weakening of their correlation. In particular, valence neutrons in nuclei with an inert core may have a large possibility to be found outside of the core. Due to the short range of the nuclear interaction, the effect of core polarization of the valence neutrons could be weakened, which results in a small effective charge and a large $M_n/M_p$ ratio. This phenomenon has been observed in light nuclei such as $^{20}$O~\cite{Jewell}, $^{15}$B~\cite{Izumi}, $^{17}$B~\cite{Ogawa}, $^{38}$S ~\cite{Kelley1997}, $^{21}$O~\cite{Heil}, $^{17,18}$O~\cite{Bernstein},  $^{20}$C~\cite{Elekes2009} and  $^{16}$C~\cite{Elekes}. %This ratio may also increase when the nucleus has an inert core~\cite{}, so the core is not easily polarized.
 Among these nuclei, $^{16}$C has  been extensively studied because an early measurement suggested a significantly reduced proton contribution compared to the neutron in the quadrupole moments of the first $2^+$ state~\cite{Elekes, Ong, Imai}. A further measurement reveals a small quadrupole polarization charge~\cite{Ong2008} but this result was revised by a recent measurement of $B(E2)$~\cite{Wiedeking}, where $M_n/M_p$ was found to be only 1.4 times its $N/Z$ value~\cite{Wiedeking, JiangY, Petri2012}.

As the neighbor of $^{16}$C, $^{15}$C is a well-known one-neutron halo nucleus, with the valence neutron occupying the $2s_{1/2}$ orbital in the ground state. Its first excited state at 0.74 MeV has a dominant single-particle configuration with a neutron in the $1d_{5/2}$ orbital and has a life time of 2.61 ns~\cite{Alburger}. Core polarization in the transition between these two states should be weakly induced due to the inert $^{14}$C core. Furthermore, the valence neutron might be further decoupled from the core due to the halo. There are some similarities in $^{15}$C and $^{16}$C with respect to their matter radius and the binding energies of the $2s_{1/2}$ and $1d_{5/2}$ single-particle orbitals~\cite{Tang}. For the $0^+_{g.s.} \rightarrow 2^+_1$ transition in $^{16}$C, some studies has attributed its valence neutron decoupling to the neutron excitation between the $2s_{1/2}$ and the $1d_{5/2}$ orbitals~\cite{Elekes2009}, since the $^{16}$C ground state has almost equal mixing of $\nu(2s_{1/2})^2$ and $\nu(1d_{5/2})^2$ configurations~\cite{Wuosmaa2010}.  Information about how the halo in $^{15}$C, induced by the $2s_{1/2}$ neutron, impacts the core polarization will enhance our understanding of the quadrupole moments of $^{16}$C mentioned above.

As the isotone of $^{15}$C, $^{17}$O has an inert $^{16}$O core and a large neutron separation energy ($S_n=4.143$ MeV). The quadrupole transition between its ground state ($5/2^+$) and first excited state ($1/2^+$) is also a single-particle transition similar to $^{15}$C. The $M_n/M_p$ value of $^{17}$O was determined to be 2.63(0.04) experimentally~\cite{Bernstein}. Comparison will be made between $^{17}$O and $^{15}$C to interpret if there is any additional valence neutron decoupling from the core due to the existence of the neutron halo in $^{15}$C. 

In this paper, we report a measurement of deuteron elastic and inelastic scattering on the one-neutron halo nucleus $^{15}$C in inverse kinematics. $M_n/M_p$, effective charges and core-polarization parameters are determined from these data, which can be used to quantify the decoupling of the valence neutron from the core. %The result also provides another constraint on the experimental discrepancy of the  $M_n/M_p$ ratio in $^{16}$C mentioned above.

\section{Experiment}

The deuteron elastic and inelastic scattering on $^{15}$C were carried out in inverse kinematics at the ATLAS in-flight facility at Argonne National Laboratory~\cite{In-flight}. The 7.1 MeV/u $^{15}$C secondary beam was produced using the neutron adding reaction on a $^{14}$C primary beam at 8 MeV/u, with an intensity of 200 particle nano Amperes (pnA). The $^{14}$C beam bombarded a 37-mm long deuterium gas cell at a pressure of 1400 mbar and temperature of 90 K. The resulting $^{15}$C beam had a rate of approximately $ 10^6$ particles per second with a negligible contamination (less than $1\%$). The secondary beam bombarded a target of either deuterated polyethylene (CD$_2)_n$ or polyethylene (CH$_2)_n$ of thickness 363\,$\mu$g/cm$^2$ and 387\,$\mu$g/cm$^2$ placed at a position along the axis of the HELIOS defined as $z = 0$\,mm. There is some proton contamination in the (CD$_2)_n$ target, so the measurement of reactions on the (CH$_2)_n$ target was used to quantify the proton content in the (CD$_2)_n$ target. The energy loss of the beam in the center of the (CD$_2)_n$ target was around 0.03 MeV/u, which has very little impact on the $Q$-value resolution or the angle uncertainty.  

\begin{figure}
  \includegraphics[width=1.0\columnwidth]{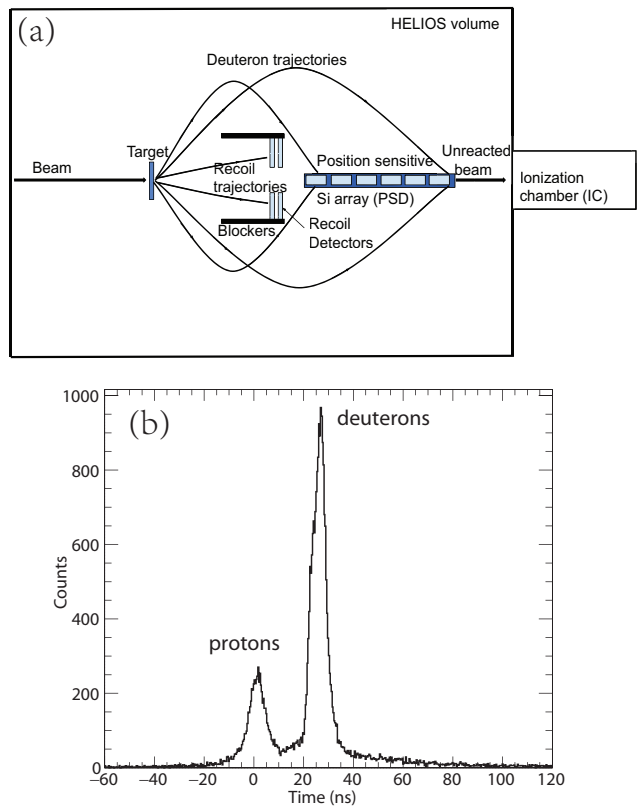}\\
  \caption{(a) A schematic of the present experimental setup. (b) The timing difference between the PSD and the recoil detectors of events generated from the reactions of $^{15}$C beams on the (CD$_2$)$_n$ target. The protons and deuterons are clearly identified as labeled. The multiple orbits were blocked by the blocker so these are the relative cyclotron period times for single orbits. }\label{protonE}
\end{figure}

\begin{figure}
  \includegraphics[width=1.0\columnwidth]{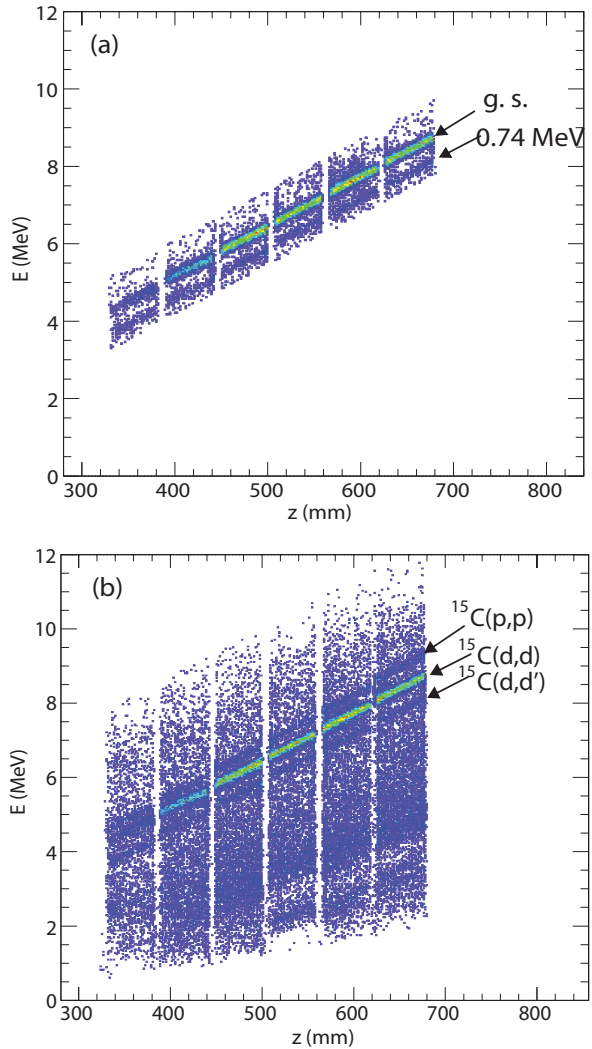}\\
  \caption{(a) Measured deuteron energies ($E$) as a function of the distance from the target ($z$) for the deuteron elastic and inelastic scattering reactions on $^{15}$C in inverse kinematics at 7.1~MeV/u with a magnetic field strength of 2.5~T. The deuteron events in (a) have a time coincidence of 20 ns with recoils. The population of unbound states is not shown here, since they are not the focus of the present discussion. Final states in $^{15}$C are labeled by their corresponding excitation energies.  (b) The measured $E$ versus $z$ spectrum of the $^{15}$C beam bombarded a (CD$_2)_n$ target by requiring a wide timing gate of 100\,ns, where the events from the reactions on the proton contamination in the (CD$_2)_n$ target can be seen. These events were utilized to deduce the proton content ratio in the (CD$_2)_n$ target. See text for details.}\label{protonE}
\end{figure}

The outgoing protons and deuterons were analyzed by the HELIOS spectrometer~\cite{WUOSMAA, LIGHTHALL} with a magnetic field strength of 2.5\,T. The silicon array, which is composed of 24 position-sensitive detectors (PSD), was placed downstream of the target covering a range of $332$\,mm $\le z \le  $ 682\,mm for the measurement of deuteron elastic/inelastic scattering (setting~1). The array was moved 60\,mm closer to the target for part of the measurement to cover smaller center of mass angles (setting~2). The spectra shown in Fig.~1-3 are for setting~1 while the differential cross sections (Fig.~4-5) includes data from both settings.  The deuterons and protons from the reactions on the (CD$_2)_n$ target were transported to the silicon array in the magnetic field. The $^{15}$C recoils were detected by $\Delta E-E$ telescopes composed of $\sim75$\,$\mu$m and $\sim$1000\,$\mu$m quadrant silicon detectors. %The experimental setup resembling that shown in Fig.~ of Ref.~\cite{}. 
In addition, the deuterons and protons traveling for more than one cyclotron period were stopped by a cylindrical plastic blocker surrounding the recoil detectors and extending in the $z$ axis. A schematic of the setup is shown in Fig.~1a.  Deuterons were identified and selected by requiring a 20-ns timing coincidence centered around the deuteron peak between a light particle detected in the PSD array and a recoil particle detected in the $\Delta E-E$ telescope (see Fig.~1b). This time gate was sufficient to discriminate the different reaction channels such as the $(d,t)$ transfer reactions or protons from fusion-evaporation reactions on the carbon in the target.

The incident beam was monitored by a fast-counting
ionization chamber (IC)~\cite{Lai} located $\sim$1000\,mm downstream
of the target. A mesh degrader was placed upstream and close to the IC to reduce the rate by a factor of $100$. The rate in the IC calculated by a discriminator threshold was used as a
scalar to count the total incident ions. The beam composition was deduced from energy-loss characters at a rate of around 15\,Hz, triggered by random coincidences in the silicon array with $\alpha$ particles from a radioactive $\alpha$ source placed in the chamber. It was found to be $>99\%$ $^{15}$C. The beam current was also checked by the elastic scattering data as discussed in the Supplemental Material~\cite{supplements,KONING,Varner}. 
%The deuterons were stopped on the PSDs and their numbers were used to determine the integrated number of incident particles times the target thickness, the luminosity. Dividing the measured experimental yield (which has been corrected for solid angle) by the calculated elastic scattering cross sections gives the luminosity of this measurement.

\section{Results}

%The deuterons detected by the PSD array correspond to $^{15}$C elastic and inelastic scattering reactions have been selected by the timing coincidence with the recoils detectors according to their cyclotron periods (see Fig.~1b). 
The energies of the deuterons ($E$) versus their detected positions ($z$) on the beam axis are plotted in Fig.~2a. Two states are clearly isolated in the spectra, which correspond to the ground state and the first excited state (0.74 MeV) of $^{15}$C. The excitation spectrum of $^{15}$C was obtained by a projection of the data  along the kinematic lines, as shown in Fig.~3. The resolution of the excitation spectrum is around 210 keV(FWHM), which was majorly contributed by the energy loss of the scattered deuterons in the (CD$_2)_n$ target and the energy/angular spread of the $^{15}$C beam. %It is clear that the relative population of the first excited state is higher for the scattering on protons than on the deuterons, which is possibly related the fact that proton has a higher sensitivity on the neutron excitation. This will be further discussed in the next section. 

There is some proton contamination in the (CD$_2)_n$ target, which needs to be subtracted from the target thickness for deducing the cross sections. The proton contamination was evidenced by the proton elastic scattering events in the $E$ versus $z$ spectrum shown in Fig.~2b, which was  confirmed by their kinematics and cyclotron period. In order to quantify the proton content, elastic scattering reactions on a (CH$_2)_n$ target were also measured. Details concentrating the proton elastic and inelastic scattering data are shown in the Supplementary Material~\cite{supplements}. The amount of proton content in the (CD$_2)_n$ was determined to be around $6\%$, which is much smaller than the uncertainty in the beam intensity (see below). This contamination was excluded by the coincidence gate discussed above and accounted for in the thickness of the (CD$_2)_n$ target. 

\begin{figure}
  \includegraphics[width=0.8\columnwidth]{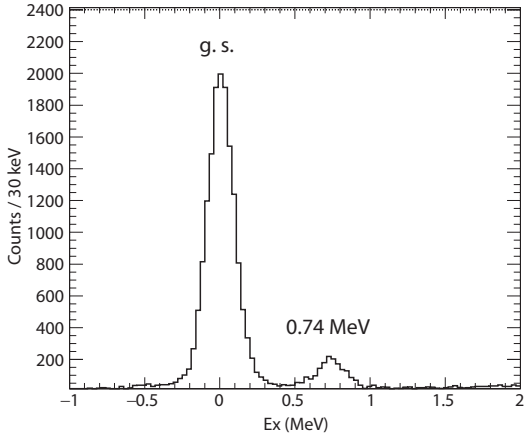}\\
  \caption{\label{fig:11BeEx} The excitation-energy spectrum of $^{15}$C bound states determined from the present measurement of $^{15}$C elastic/inelastic scattering  on deuterons. The ground state and first excited state are separated as labeled. Same gates as Fig.~2a were used. }
\end{figure}

%\linespread{1.3}
%\begin{table}
%\caption{\label{tab:expt-S} (delete) Normalization factors, deformation lengths fitted to the experimental data using the DA1p~\cite{Zhang} as the starting point. The fit was optimised to simultaneously reproduce the elastic and inelastic scattering data. The uncertainties resulted from the fitting are shown in the parentheses.  }
%\begin{ruledtabular}
%  \begin{threeparttable}
%\begin{tabular}{ccc}
%\textrm{$\lambda_R$}&
%\textrm{$\lambda_I$}&
%%\textrm{$\chi^2_{el}/n$}&
%\textrm{$\chi^2_{inel}/n$}&
%\textrm{$\delta$(fm)}\\
%\colrule\\
%  0.939(2)   & 0.721(4) &    1.04(1) \\
%   \end{tabular}
%    \end{threeparttable}
%\end{ruledtabular}
%\end{table}

%\section{Angular distributions}

\begin{figure}[htp]
  \includegraphics[width=1.0\columnwidth]{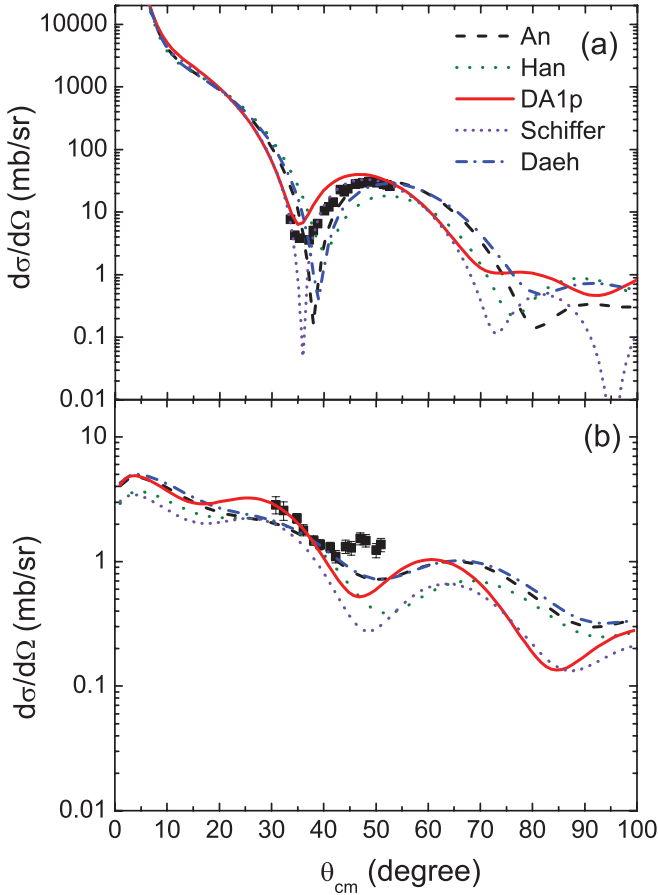}\\
  \caption{Experimental (black points) and calculated (lines) differential cross sections for the deuteron elastic scattering (a) and  inelastic scattering to the first excited state of $^{15}$C (b).  Corresponding  calculation results using different OPs are labeled in the figure. A deformation parameter of $\beta_d =0.29$ was used for the inelastic scattering differential cross sections. }
\end{figure}

The differential cross sections of deuteron elastic scattering and inelastic scattering excitation to the first excited state of $^{15}$C were deduced from the present data. %The two data points of the smallest center of mass (c.m.) angles were obtained by moving the array 6 cm upstream. 
Each PSD was divided into two or three bins as statistics allowed. Center-of-mass angles and solid angles were determined from the detector geometry and kinematics of each state, where the uncertainties were governed by the position of the silicon array. There was a maximum uncertainty of around 4 mm in the position of the silicon array, which result in an uncertainty of around $0.3^{\circ}$ in the center-of-mass angles.  The total beam exposure was determined from the IC counts. It was checked by  comparing the $^{15}$C$+p$ elastic scattering data to the calculated cross sections (see Supplementary Material~\cite{supplements}) and also to that of $^{12,13}$C$+p$ at the same incident beam energy taken from Ref.~\cite{Guratzsch}. The absolute differential cross sections are plotted in Fig.~4. The error bars are statistical only. There is an overall systematic uncertainty estimated to be around 20$\%$, primarily driven  by the uncertainties in the beam intensity and in the target thickness. 

The measured elastic scattering angular distributions were analyzed in the framework of the optical potential models. The optical potential (OP) consists of real, imaginary and spin-orbit component of a Woods-Saxon form as below,
\begin{equation}\begin{aligned}
V(r)=&-V_{0} f\left(x_{0}\right)-i W_{v} f\left(x_{v}\right)+4 W_{s} \frac{d f\left(x_{d}\right)}{d x_{d}} \\
&-V_{s o}\left(\frac{\hbar}{m_{\pi} c}\right)^{2} \frac{1}{r} \frac{d f\left(x_{s o}\right)}{d r}(\vec{L} \cdot \vec{s}),
\end{aligned}\end{equation}
where
\begin{equation}\begin{array}{l}
f\left(x_{i}\right)=1 /\left[1+\exp \left(x_{i}\right)\right] \\
x_{i}=\left(r-r_{i} A^{1 / 3}\right) / a_{i}, \quad i=0, v, s \text { and } s o .
\end{array}\end{equation}
Global OPs from An {\it et al.}~\cite{An}, Han {\it et al.}~\cite{Han}, Daehnick {\it et al.}~\cite{Daehnick}, Schiffer {\it et al.}~\cite{Schiffer} and DA1p~\cite{Zhang} were used for the deuteron elastic scattering data. The global optical model potentials of  DA1p\cite{Zhang} best reproduce the experimental elastic scattering cross sections. These potentials were developed by fitting the experimental data of the 1p-shell nuclei, so were explicitly derived for nuclei in this mass range. 

The matter deformation parameter $\beta_d$ is extracted by normalizing the theoretical calculations to the inelastic scattering differential cross sections using the relationship $(d\sigma/d\Omega)_{exp}= \beta_d^2(d\sigma/d\Omega)_{DWBA}$. This can be related to the deformation length, $\delta_d$, by the relation $\beta_d = \delta_d/(r_0 A^{1/3})$. The parameter $r_0$ is taken to be 1.2\,fm. For the inelastic scattering data, one-step distorted wave Born approximation framework was applied using two computer codes, FRESCO~\cite{fresco} and PTOLEMY~\cite{Macfarlane1978}. These two codes yield results within $2\%$ for the same OP parameters which has no impact on the present analysis.  A deformation parameter of $\beta_d = 0.29(3)$ was extracted using the DA1p potential of Ref.~\cite{Zhang}.  The uncertainties resulting from  different OPs were also investigated. In Fig.~4b, the same  $\beta_d = 0.29$  was used in the calculations for  inelastic scattering cross sections using different global OPs. The dependence on the OP parameters is acceptable at around 25$^{\circ}$ to 45$^{\circ}$ in the center of mass frame. %Therefore, these data were given a larger weight in investigating the matter deformation length of $^{15}$C. %Furthermore, for the proton inelastic scattering, the sensitivity on the proton and neutron have a large dependence on the reaction energies, and there is not enough experimental or theoretical study  at this low incident energy.  
%Therefore, there will be large uncertainties in the Mn/Mp value if 15C proton inelastic scattering data were used. Here
%Therefore, only $^{15}$C$+d$ inelastic scattering data will be used to investigate the deformation length of $^{15}$C. 
The calculation is less successful at larger angles where  core excitation, continuum coupling~\cite{CHEN2016,CHEN2016Mar} or three/four-body effects~\cite{Descouvemont, Descouvemont2017} may be expected to impact the cross sections. %Future study about these effects in the deuteron scattering on halo nucleus $^{15}$C is expected to explain this discrepancy.

\begin{figure}[htp]
  \includegraphics[width=1.0\columnwidth]{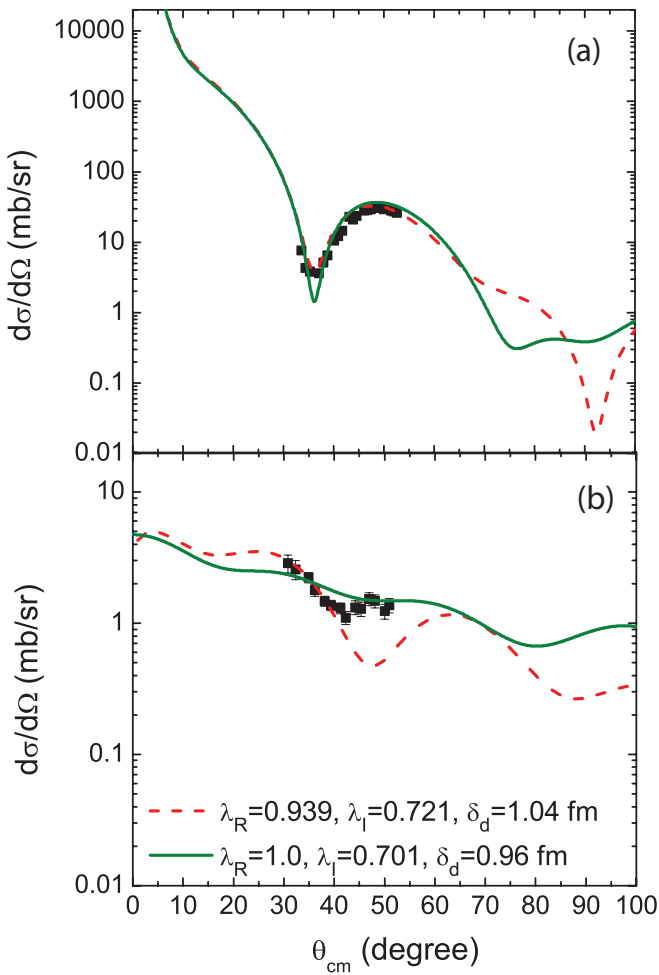}\\
  \caption{Experimental (black points) and calculated (lines) differential cross sections for  deuteron elastic scattering and  inelastic scattering to the first excited state of $^{15}$C. The OP of DA1p was normalized by the listed factors. Deformation length is also listed which was used in the calculation of the  inelastic differential scattering cross sections. See text for details. }
\end{figure}

Coupled channel calculations were performed using the code FRESCO in the framework of the rotational model to extract the deformation length ($\delta_d$) from the  deuteron inelastic scattering data. % 9Li paper
It is known that for some weakly-bound nuclei, the OP parameters need to be normalized to best fit the data~\cite{CHEN2016,CHEN2016Mar}. Therefore, the normalization factor $\lambda _R$ and $\lambda_I$ were applied on the well depth of the real ($V_0$ in eq.~1) and imaginary part ($W_s$ and $W_v$ in eq.~1 ). DA1p global OP parameters were used as the starting point of the fit. These normalization factors were obtained by performing a searching process with the code SFRESCO~\cite{fresco}, where the $\chi^2$ minimization method was used. %These normalized OPs were used as the start points to search for the best fit of the inelastic scattering.
 The deformation length was also searched using SFRESCO, together with the normalization factors $\lambda _R$ and $\lambda_I$ being varied. $\lambda _R=0.939(2)$ and $\lambda_I=0.721(4)$ were obtained. The fitting of these two parameters were mostly guided by the elastic scattering data which has much larger cross sections. 
The deformation length of $\delta_d = 1.04(11)$\,fm was extracted by fitting to the inelastic scattering differential cross sections. The fitting uncertainty is much smaller than the systematic and statistic uncertainties. Overall, the uncertainty is around 10$\%$. This leads to a deformation parameter of $\beta_d =  0.35(4)$, which agrees with the value $\beta_d = 0.29(3)$ obtained using global OPs if $\lambda _R$ and $\lambda_I$ were not applied (see Fig.~4 red dashed lines). %The uncertainty coming from the data statistic and systematic uncertainty is around 10$\%$. %A variety of  optical model potentials~\cite{Han,An,Daehnick,Varner,Lohr} have been applied to the data to search for the deformation length and $10\%$ uncertainty was obtained. 

The coupled channel effect can be embedded in the imaginary potential of the OM potential. Therefore, we have performed the DWBA calculation with the search of the OM potential by fixing the real potential ($\lambda_R$=1.0) but varying the imaginary potential ($\lambda_I$). A normalization factor of $\lambda_I=0.701$ was obtained and the calculated result was shown in Fig.~5 (green solid lines). The resulted deformation length agrees with the couple channel calculation within uncertainties. It worth noting that the resulted normalization factor of the imaginary potential comes not just from the coupled channel effect, but may also because the existing systematic OM potential did not include this weakly bound nucleus $^{15}$C. 

The proton and neutron quadrupole matrix element $M_{n(p)}=\left\langle J_f\left\|\sum_{n(p)} r^{2} Y_{2}\right\| J_i\right\rangle$ can be obtained by comparison of measurements of a transition using two experimental probes, which have different sensitivities to neutron and proton contributions. Because the relative electric quadrupole transition probability $B(E2)$ is in principle a purely electromagnetic probe, $M_p$ is determined by 
\begin{equation}B\left(E 2, J_i^{+} \rightarrow g.s.\right)=e^{2} \frac{M_{p}^{2}}{2 J_{i}+1},\end{equation}
and the Coulomb deformation length $\delta_{p}$ is related to $M_p$ by~\cite{Bernstein1983}
\begin{equation}\delta_{p}=\frac{4 \pi}{3 Z R} M_{p}.\end{equation}
The nuclear radius $R$ is taken as the
standard value of $1.2A^{1/3}$\,fm~\cite{IWASAKI20007,Elekes2009,Kondo}. The $B(E2)$ value, proton quadrupole matrix element and proton deformation length of $^{15}$C are determined to be 0.97(2)\,e$^2$fm$^4$, $M_p=2.42(3)$\,fm$^2$ and $\delta_p=0.57(1)$\,fm, respectively, from the lifetime measurement of the $5/2^+$ state~\cite{Alburger}. 

For the other experimental probe, we will use the present deuteron inelastic-scattering data. The deformation length $\delta_d$ is associated with $M_n/M_p$ and the interaction strength of neutron and proton $b_n/b_p$ in the following relationship~\cite{Bernstein1983,Bernstein1981}:
\begin{equation}\frac{\delta_{d}}{\delta_{p}}=\frac{1+\left(b_{n} / b_{p}\right)\left(M_{n} / M_{p}\right)}{1+\left(b_{n} / b_{p}\right)(N / Z)}.\end{equation}
For different probes, $b_n/b_p$ varies substantially. For example, $b_n/b_p$ for protons changes from 0.8 to 3 with different incident energies and the  value at incident energies less than $10$ MeV is still not well studied theoretically or experimentally~\cite{Bernstein1981,Kanada}. However, for the deuteron, $b_n/b_p$ is always equal to one and has very little dependence on the energies because of its isoscalar nature~\cite{Bernstein1983,Bernstein1981}. % Therefore, the $(d, d')$ scattering is utilized to deduce $M_n$ rather than the $(p, p')$ scattering. %This is another reason for deducing the deformation length from the deuteron inelastic scattering data. 

%If we look back at the higher population of the $5/2^+$ state in the proton inelastic scattering on $^{15}$C, it may be due to the higher sensitivity of protons on neutron dradrupole excitations. But the qualitatively, $b_n/b_p=3$ for the proton probe cannot fully explain the difference and this question requires further study, which is beyond the scope of the present work.  Also OP

With $b_n/b_p=1$ and deformation length of $\delta_d=1.04(11)$\,fm, according to equation (5), $M_n/M_p =3.6(4)$ was determined for $^{15}$C. This value is much larger than the $N/Z$ value of $^{15}$C, with a ratio $M_n/M_p/(N/Z) = 2.4(3)$. % More detailed discussion is needed to interpret its core decoupling effect. 

In the traditional core + valence shell model framework, $M_n$ and $M_p$ are described in terms of valence space quadrupole matrix elements $M_n^{\prime}$, $M_p^{\prime}$ and core polarization parameters ($ \Delta^{p n}$,  $\Delta^{n n}$,  $\Delta^{p p}$). $M_n^{\prime}$ and $M_p^{\prime}$ are the static quadrupole moments and can be deduced from the shell structure of the valence neutrons or protons. $\Delta^{x y}$ is the core-polarization parameter corresponding to core ($x$) polarization by valence nucleon ($y$), which reflects the amount of core polarization per unit of contribution from the valence nucleon~\cite{Jewell}. The neutron and proton quadrupole matrix element is calculated using the following equation~\cite{Bernstein1983}, 
\begin{equation}
M_{n}=M_{n}^{\prime}\left(1+\Delta^{n n}\right)+M_{p}^{\prime} \Delta^{n p},
\end{equation}
and
\begin{equation}
M_{p}=M_{n}^{\prime} \Delta^{p n}+M_{p}^{\prime}\left(1+\Delta^{p p}\right).
\end{equation}
With neutron and proton effective charge represented by 
\begin{equation}e_{n}=\Delta^{p n} \text { and } e_{p}=1+\Delta^{p p}.\end{equation}
If we consider the $^{14}$C core to be inert, and $M_{p}^{\prime}$ is close to zero,  we have
\begin{equation}
M_{n}=M_{n}^{\prime}\left(1+\Delta^{n n}\right),
\end{equation}
and
\begin{equation}
M_{p}=M_{n}^{\prime} \Delta^{p n}=M_{n}^{\prime} e_{n}.
\end{equation}

For $^{15}$C the $1/2^+$ to $5/2^+$ transition is dominated by the
neutron $1d_{5/2}$-$2s_{1/2}$ one-body transition density. The shell model calculation using the YSOX interaction~\cite{Yuan} predicts $M_{n}^{\prime}$=5.1872\,fm$^2$ and $M_{p}^{\prime}$=0.4433\,fm$^2$. From this result, it is reasonable to assume $M_{p}^{\prime} \sim 0$ compared to $M_{n}^{\prime}$. In Ref.~\cite{Wiedeking}, $M_{n}^{\prime}$ is calculated to be 6.0\,fm$^2$ using harmonic oscillator wave functions, also in agreement with this shell model prediction. Using these values, $e_n=\Delta^{p n}=0.4\sim0.46$ and $\Delta^{n n}=0.4\sim0.6$ are ranges determined from the present measurement. These values agree with the empirical values~\cite{Brown}.  %%citation....Discussion

\begin{table*}
	\caption{Calculated and experimental ground state energies E\textsubscript{g.s.}, excitation energies of the first $\frac{5}{2}$\textsuperscript{+} state E\textsubscript{x}, E2 transitions B(E2), M1  moments and the $M_n/M_p$ values of \textsuperscript{15}C. The first three columns correspond to results from {\it ab-initio} NCCI calculations  with Daejeon16 at  $\hbar\Omega$ =17 MeV within different basis spaces, while the next three columns correspond to results at  $\hbar\Omega$ = 18 MeV. The extrapolated NCCI results along with the experiments are shown for comparison. See the text for details. }
	\resizebox{\textwidth}{!}{
	\begin{tabular}{ccccccccc}
		\hline 
		\hline 
		\\
		\centering  
		$\hbar\Omega$ & \space  & 17 MeV  &\space  &\space & 18 MeV &\space &Extrapolation & Experiment \\
		 \cline{2-5}  \cline{6-7}& \space \\
		N\textsubscript{max} & 4&6&8 & 4&6&8 & \space& \\
		\hline 
		\\
		E\textsubscript{g.s.}($\frac{1}{2}$\textsubscript{1}\textsuperscript{+}) [MeV] & -100.034
   &  -104.146 &-106.091
   & -100.194
   &   -104.134 & -106.019 & -107.793(45)&  -106.503~\cite{nndc}  \\
		\\
		E\textsubscript{x}($\frac{5}{2}$\textsubscript{1}\textsuperscript{+}) [MeV]  &  0.556 & 0.941 & 1.169 
   & 0.494 & 0.908 & 1.148& 1.440(9) & 0.740(15)~\cite{nndc}   \\
		\\
		B(E2;$\frac{5}{2}$\textsubscript{1}\textsuperscript{+}\textrightarrow$\frac{1}{2}$\textsubscript{1}\textsuperscript{+}) [e\textsuperscript{2}fm\textsuperscript{4}]  &  0.699 & 0.938 & 1.115 
   & 0.658 & 0.899 & 1.080 & 2.025(30) & 0.97(2)~\cite{Alburger}     \\
		\\
		$\mu$\textsubscript{g.s.}($\frac{1}{2}$\textsubscript{1}\textsuperscript{+}) [$\mu$\textsubscript{N}] & -1.723
   &  -1.717 &-1.711
   & -1.720
   &  -1.714 & -1.709 & -1.633(53)
   &  $|1.315(70)|$~\cite{Raghavan}  \\
   \\
   $\mu$($\frac{5}{2}$\textsubscript{1}\textsuperscript{+}) [$\mu$\textsubscript{N}]& -1.464
   &  -1.442 &-1.428
   & -1.467
   &   -1.443 & -1.429 & -1.407(9)
   &  -1.758(30)~\cite{Raghavan}  \\
   \\
   M\textsubscript{n}/M\textsubscript{p} ($\frac{5}{2}$\textsubscript{1}\textsuperscript{+}\textrightarrow$\frac{1}{2}$\textsubscript{1}\textsuperscript{+}) & 3.870
   & 3.649
   & 3.578
   & 3.876
   &  3.652
   & 3.575
   & 3.529(6)
   &  3.6(4)  \\
		\hline
		\hline
		
	\end{tabular}}
	\centering
    
	\end{table*}

\section{Discussion}

\subsection{Comparison with $^{17}$O and other C isotopes}
%read 16C paper
A comparison can be made to the $N=9$ isotone $^{17}$O. The B(E2) value of $^{17}$O between the g.s. and first excited state was measured to be 1.036\,e$^2$fm$^4$. $M_p=2.54$\,fm$^2$ and $e_n=0.42$ were deduced accordingly~\cite{Bernstein}. With the B(E2) value of its mirror nucleus $^{17}$F corrected by the Coulomb correction factor, $M_n/M_p=2.63(0.04)$ was deduced for $^{17}$O~\cite{Bernstein}. This value is 2.34 times larger than its $N/Z$ value. Considering the large $S_n$ of $^{17}$O, its large $M_n/M_p$ is primarily due to the valence neutron decoupling from the inert $^{16}$O core. For $^{15}$C, the ratio between $M_n/M_p$ and $N/Z$ is 2.4(3), which is close to $^{17}$O. This indicates that the degree of neutron decoupling of $^{15}$C and $^{17}$O is similar and there is no clear additional reduction of core polarization in the halo nucleus $^{15}$C. This suggests that the core decoupling effect induced by the halo is not prominent in $^{15}$C. %It is noted that the $M_n/M_p/(N/Z)$ value of nuclei with one valence neutron, like $^{17}$O and $^{15}$C is usually larger than in even-even nuclei (for example $^{18}$O, $^{38}$S~\cite{Kelley1997} and $^{32}$Si~\cite{Cottle}), so it is more appropriate to compare $^{15}$C to its isotone rather than other even-even nuclei, in order to interpret the effect of the halo.

Neutron-rich carbon isotopes have attracted much attention with regards to the question of spatially extended and decoupled valence neutrons. The present result provides another insight into this discussion. Since the  halo nucleus $^{15}$C does not show a strong core decoupling effect compared to $^{17}$O, considering the similarities in $^{15}$C and $^{16}$C~\cite{Tang}, one may expect that the ratio between $M_n/M_p$ and $N/Z$ would be similar in $^{18}$O and $^{16}$C. %which agrees with the result in Ref..  
For $^{16}$C, the latest measurement of the ratio $M_n/M_p/(N/Z)$ is 1.4~\cite{Wiedeking}, similarly to $M_n/M_p/(N/Z)=1.8$ in $^{18}$O. In Ref.~\cite{Wuosmaa2010}, it was found that $^{16}$C may be described without invoking very exotic phenomena, which is also in line with the present interpretation. For the more neutron rich carbon isotopes, there are also recent studies showing that no evidence was found for
dramatic changes in the behavior of the $B(E2)$  up to $^{20}$C~\cite{Petri}. %Future measurements which determine their neutron and proton contributions are still required to further understand the decoupling effect along the carbon isotopic chain. 

It is worth noting that the matter radius of $^{15}$C~\cite{Kanungo} is only moderately increased due to the halo compared to other carbon isotopes, indicating that the halo structure in  $^{15}$C may not be
sufficiently pronounced to cause a  strong decoupling of the valence neutron. For instance, the matter radii of $^{14}$C, $^{15}$C and $^{16}$C are 2.33(7), 2.54(4) and  2.74(3)\,fm~\cite{Kanungo}, respectively, which have been well reproduced by the calculation with a Woods-Saxon potential considering their single-particle configuration in the $2s_{1/2}$ and $1d_{5/2}$ orbitals~\cite{Tang}. %The $^{16}$C radius is a little larger because the ground state has almost equal mixing of $\nu(2s_{1/2})^2$ and $\nu(1d_{5/2})^2$ configurations~\cite{Wuosmaa2010}, where  
The valence neutron in $^{15}$C is polarizing the core to an extent similar to its isotones  without a halo structure. This conclusion is similar to that in Ref.~\cite{Tang}, where no special theoretical treatment for the neutron halo is needed to explain various experimental results of $^{15}$C. %and the 
Experimental studies of other typical halo nuclei with larger matter radii, for example, $^{19}$C~\cite{Kanungo}, $^{22}$C~\cite{Tanaka} or $^{11}$Li~\cite{Al-Khalili}, will be interesting in order to further understand the core-polarization effect in the halo nuclei. 

\subsection{{\it Ab-initio} calculation}

{\it Ab-initio} no-core configuration interaction (NCCI)~\cite{Barrett2013,Maris:2008ax,Navratil:2000ww}  calculations for $^{15}$C were carried out with the Daejeon16 interaction~\cite{Shirokov:2016ead}. Using the MFDn code ~\cite{Aktulga,MARIS201097}, we diagonalized the Hamiltonian of the system in a harmonic oscillator basis  which is characterized by the basis energy scale $\hbar\Omega$ and the basis truncation parameter
N\textsubscript{max}(defined as the maximum of the total oscillator quanta above the minimum for $^{15}$C that satisfies the Pauli principle).   We summarize the NCCI results in Table I.  A simple 3-point exponential extrapolation~\cite{Maris:2008ax, Maris:2019etr} was used for the ground state and first excited state energies, as well as the magnetic dipole (M1) moments at two $\hbar\Omega$ values. For the extrapolation of B(E2), we adopt the extrapolation formula for electric quadruple
transitions in Ref.~\cite{Odell}. The extrapolations of observables in Table I were obtained from the average of extrapolated values at  $\hbar\Omega$ = 17 MeV and 18 MeV, which are values approximating the variational minimum of the ground state energy in the largest basis space. 

The ground state energy of $^{15}$C is in reasonable agreement with experimental data. 
The extrapolated excitation energy of the first excited state is about 0.72 MeV higher than experiment, which is moderately acceptable. The extrapolated B(E2) value of the NCCI calculation is about two times the experimental result. E2 transition matrix elements are very sensitive to the long range tails of the nuclear wave function, which is not adequately accommodated in the limited harmonic oscillator basis. Ref.~\cite{Caprio} reported that the ratio B(E2)/(e$^{2}$r$_{p}^4$) (r$_{p}$ is the r.m.s. point-proton radius) exhibits good convergence in $^{7}$Li and $^{10}$Be with both N\textsubscript{max} and $\hbar\Omega$.  We test the convergence of this ratio with respect to N\textsubscript{max} and $\hbar\Omega$.  We find that B(E2)/(e$^{2}$r$_{p}^4$) in $^{15}$C does not have as good convergence with respect to N\textsubscript{max} as that in $^{7}$Li and $^{10}$Be. The M1 moments of the ground state and first excited state are in reasonable agreement with the experimental results considering that contributions of two-body currents are not included. Remarkably, the calculated $M_n/M_p$ value appears to be well-converged and agrees well with the present experimental result.  This suggests that $M_n/M_p$ may provide a robust ratio for comparing experiment with theory along similar lines or reasoning as the ratios presented in Ref.~\cite{Caprio, Caprio:2021umc}. 

The NCCI  approach provides an overall reasonable description of the $^{15}$C bound states and supports the finding that the core-decoupling effect is not remarkable in $^{15}$C. We note that the Daejeon16 interaction was obtained by using phase-equivalent transformations(PETs) to adjust off-shell properties of the similarity renormalization group evolved chiral effective field theory NN interaction to fit selected binding energies and spectra of p-shell nuclei in an {\it ab-initio} approach~\cite{Shirokov:2016ead}. Therefore we may anticipate improving the Daejeon16 interaction in the future by fitting properties of sd-shell nuclei in order to  describe better the excitation energy of the first excited state and the B(E2) of $^{15}$C.  %Further calculation for other carbon isotopes will be presented in another theoretical paper.

\section{Summary}
Deuteron elastic scattering on $^{15}$C and inelastic scattering to the first excited state of $^{15}$C have been studied in inverse kinematics with the HELIOS spectrometer. Matter deformation has been determined from the deuteron inelastic scattering data. The ratio of neutron and proton quadruple matrix elements $M_n/M_p/(N/Z)= 2.4(3)$ and effective changes deduced from the deformation length indicate that relative to $^{17}$O, there no evidence for additional  decoupling of the valence neutron induced by the halo in $^{15}$C. The NCCI calculation with the Daejeon16 interaction provides an overall reasonable description of the properties of the two lowest bound states of $^{15}$C. The result  supports an overall modest valence-neutron decoupling picture in the neutron-rich carbon isotopes.

\begin{acknowledgments}

The authors would like to acknowledge the efforts of the support and
operations staff at ATLAS.  This research used resources of Argonne National Laboratory's ATLAS
facility, which is a Department of Energy Office of Science User Facility. This material is based upon work supported by the U.S. Department of Energy, Office of Science, Office of Nuclear Physics, under Contract No.~DE-AC02-06CH11357
(ANL), No. DE-FG02-87ER40371, DE-SC0014552(UCONN) and No. DE-SC0018223 (SciDAC-4/NUCLEI). D. K. S acknowledges U.K. Science and Technology Facilities Council [Grants No. ST/P004423/1 and ST/T004797/1]. 
%Grant Nos.~DE-FG02-96ER40978 (LSU), DE-FG02-95ER-40934 (ND), DE-SC0014552 (UConn), DE-AC02-05CH11231(LBNL) and DE-SC0009971 (CUSTIPEN). C.~X.~Y. and Y.~L.~Y acknowledges the National Natural Science Foundation of China 11775316, 11535004, 11875074. This research used computational
%resources of the National Energy Research Scientific Computing Center (NERSC), a U.S.~Department of Energy, Office of Science, user facility supported under Contract~DE-AC02-05CH11231, and of the Argonne Laboratory Computing Resource
%Center; support from the NUCLEI SciDAC program is gratefully acknowledged. 
H. L. and X. Z are supported by the Natural Science Foundation of Gansu Province, China, Grant No. 20JR10RA067 and by the Central Funds Guiding the Local Science and Technology Development of Gansu Province. This research used the computing resources of National Computing Center in Jinan and Gansu Advanced Computing Center.
We gratefully acknowledge use of the Bebop cluster in the Laboratory Computing Resource Center at Argonne National Laboratory. Data associated with this experiment can be obtained by reasonable request to the author.
\end{acknowledgments}

\bibliography{15Cdd}

\end{document}